\documentclass[fleqn,usenatbib]{mnras}

\usepackage[T1]{fontenc}
\usepackage{newtxtext,newtxmath}
\usepackage{graphicx}
\usepackage{amsmath}
\usepackage{xcolor}
\usepackage{textgreek}
\usepackage[utf8]{inputenc}
\usepackage[english]{babel}
\usepackage{color,colortbl}
\usepackage{tensind}
\tensordelimiter{?}
\DeclareGraphicsExtensions{.bmp,.png,.jpg,.pdf}
\usepackage{verbatim}
\usepackage[normalem]{ulem}
\usepackage{orcidlink}
\usepackage{soul}

\definecolor{forestgreen}{rgb}{0.13, 0.55, 0.13}
\definecolor{linkcolor}{rgb}{0.0,0.3,0.5}

\graphicspath{{./figs/}}

\title[Galaxy and halo root systems]{Galaxy and Halo Root Systems: Fingerprints of Mass Assembly}

\author[Neyrinck, Arag\'on-Calvo \& Szapudi]{%
Mark Neyrinck\orcidlink{0000-0002-2618-5790}$^{1,2}$\thanks{E-mail: neyrinck@hawaii.edu},
Miguel Arag\'on-Calvo$^{3}$\thanks{E-mail: maragon@astro.unam.mx},
and Istv\'an Szapudi\orcidlink{0000-0003-2274-0301}$^{1}$\thanks{E-mail: szapudi@hawaii.edu}\\
$^{1}$Institute for Astronomy, University of Hawai\`i at M\=anoa, 2680 Woodlawn Drive, Honolulu, HI 96822, USA\\
$^{2}$Blue Marble Space, 600 1st Avenue, 1st Floor, Seattle, Washington 981049, USA\\
$^{3}$Universidad Nacional Autónoma de México, Instituto de Astronomía, A.P. 106, 22800. Ensenada, B.C., México
}

\date{Accepted XXX. Received YYY; in original form ZZZ}
\pubyear{2026}

\begin{document}
\label{firstpage}
\pagerange{\pageref{firstpage}--\pageref{lastpage}}
\maketitle

\begin{abstract}
We discuss what we call halo or galaxy root systems, collections of particle pathlines that show the infall of matter from the initial uniform distribution into a collapsed structure. The matter clumps as it falls in; projected through time, it produces filamentary density enhancements analogous to tree roots and branches, or branching river networks. This relates to the larger-scale cosmic web, but is defined locally about one of its nodes: a physical, geometric version of a merger tree. We find dark-matter-halo root systems on average to exhibit more roots and root branches for the largest cluster haloes than for small haloes, indicating their more complicated assembly, even in dark matter without gas physics. Also, we find that high spin manifests as curvier roots, and that many have mass contributions that start far away from central, compact protohalo cores. Root systems also show sensitivity to anisotropic infall, which we see some evidence for in a simulation with rather low box size.
\end{abstract}


\section{Introduction}
\label{sec:intro}

The cosmic web is the largest flow network in the Universe, the branching network of matter that connects galaxies on intergalactic scales. Branching networks with similarities to the cosmic web abound on much smaller scales in nature too, visible on Earth in more familiar systems, such as river networks; human infrastructure networks; trees, roots, and mycelial networks; underground flow and fracture networks; and circulatory, respiratory, and neural systems in our bodies.

There is some understanding of why these systems in many cases look similar, and even artistic modes of knowing have aided in this; for example, room-sized art installations by Tom\'{a}s Saraceno based on spiderwebs \citep{Ball2017} inspired the finding that the cosmic web is a structural-engineering spiderweb (a network of links that can be all in tension, or all in compression) \citep{NeyrinckEtal2018}. \citet{DiemerFacio2017} have also explored constructing textile representations of the cosmic web. Some networks of neurons in the brain look similar to the cosmic web, as well \citep{VazzaFeletti2020}. Some artistic efforts have encouraged understanding of that correspondence \citep[e.g.][]{NeyrinckEtal2020}, including a tentative finding that local cosmic webs, just like neurons \citep{CuntzEtal2010}, and urban transportation networks \citep{leite2024similarity}, balance the efficiency of minimal-spanning trees with as-direct-as-possible transport to a single hub. Measurements of branching systems are obviously of use for medical-imaging diagnosis; for example, the fractal dimension of the bronchial lung structure is of diagnostic use \citep{LungFractalDim}. Finding evidence of both differences and similarities between systems can help to understand how both work.

One rigorous similarity between trees and cosmic webs is already known: trees are often structural spiderwebs (in equilibrium, if entirely in either tension or compression), and therefore share geometric underpinnings with the cosmic web. But a difference between a tree and the cosmic web is that trees are compact, with a single trunk. The cosmic web, on the other hand, is in principle an infinite, percolating structure (if observed on a constant, recent timeslice). Locally though, cosmic-web patches around haloes can resemble trees. In particular, this resemblance comes out if we look at the accretion pattern that underlies a halo.

We call the web formed by matter as it collapsed in the past into a halo or galaxy a {\it root system}. In this introductory work, we just analyze haloes, but also touch on what we expect for galaxy root systems. A root system is a projection of a halo through time, instead of space; it is the projected ensemble of matter trajectories, or pathlines, for all particles that end up in the halo. Pathlines are common concepts in fluid mechanics. But a couple of choices make the pattern they form particularly interesting, to our knowledge not explored before, cosmologically: (1) we view it in comoving coordinates, so infall dominates other motions; and (2) we subtract off any halo bulk motions that might obscure the infall pattern; we keep the centroid of the protohalo that eventually forms the halo in the same place at each snapshot that we use to render it.

An essential concept in galaxy formation is a galaxy or halo merger tree \citep[e.g.][]{SrisawatEtal2013}. This can help to understand protogalaxy and protocluster assembly, and star-formation histories \citep[e.g.][]{ChiangEtal2013,ChiangEtal2017}. But `tree' in `merger tree' refers to the abstract computer-science concept of nodes in a hierarchy; root systems, on the other hand, are physical geometric branching structures. 

There is also a link between root systems and `watershed superclusters' or `basins of attraction,' such as Laniakea \citep{TullyEtal2014,DupuyCourtois2023,Valade2024}, defined by dominantly inward flows. A watershed supercluster might be defined as something that eventually would form a giant halo, at least without the structure-freezing effect of cosmic acceleration \citep{Araya-MeloEtal2009}. Streamlines of reconstructions of the velocity field in such a region \citep[e.g.][]{HoffmanEtal2024} would approximate that halo's root system. However, for this comparison, it is important to note that the large-scale velocity flows usually used to demarcate watershed superclusters are defined with streamlines (curves tangent to the velocity field at a fixed time), not pathlines (actual paths that matter has taken in the past), which is what we investigate here.

As we indicate with `fingerprints of mass assembly' in the title, root systems record tracks of mass assembling into haloes. They should be sensitive to halo ellipticity and connectivity; these are known to correlate with accretion history \cite{GouinEtal2021}. Halo substructure and connectivity, possible to infer observationally \citep[e.g.][]{EinastoEtal2025} as well as in simulations, may leave imprints both inside and outside a root system.

So, we also expect features in them to encode much (possibly all?) information relevant to assembly bias \citep[e.g.][]{GaoWhite2007}, i.e., differences in the clustering of halo populations of the same final mass, arising from different formation histories, spins, and types of progenitors.

This paper is structured as follows. In the next subsection, we visually introduce the concept with a two-dimensional example. After discussion of the method, we give several types of results: a discussion of the visual impression of the root systems; possible links to star formation; resolution effects; how to use root systems to study an effect such as the level of infall isotropy; and root-system shape. Then, there is a short section about the capability of root systems to carry information about dark-matter haloes and galaxies. The final section is a conclusion, including some speculation about root systems' broader relevance.

\subsection{A 2D Illustration}
\begin{figure*}
        \centering
    	\includegraphics[width=\textwidth]{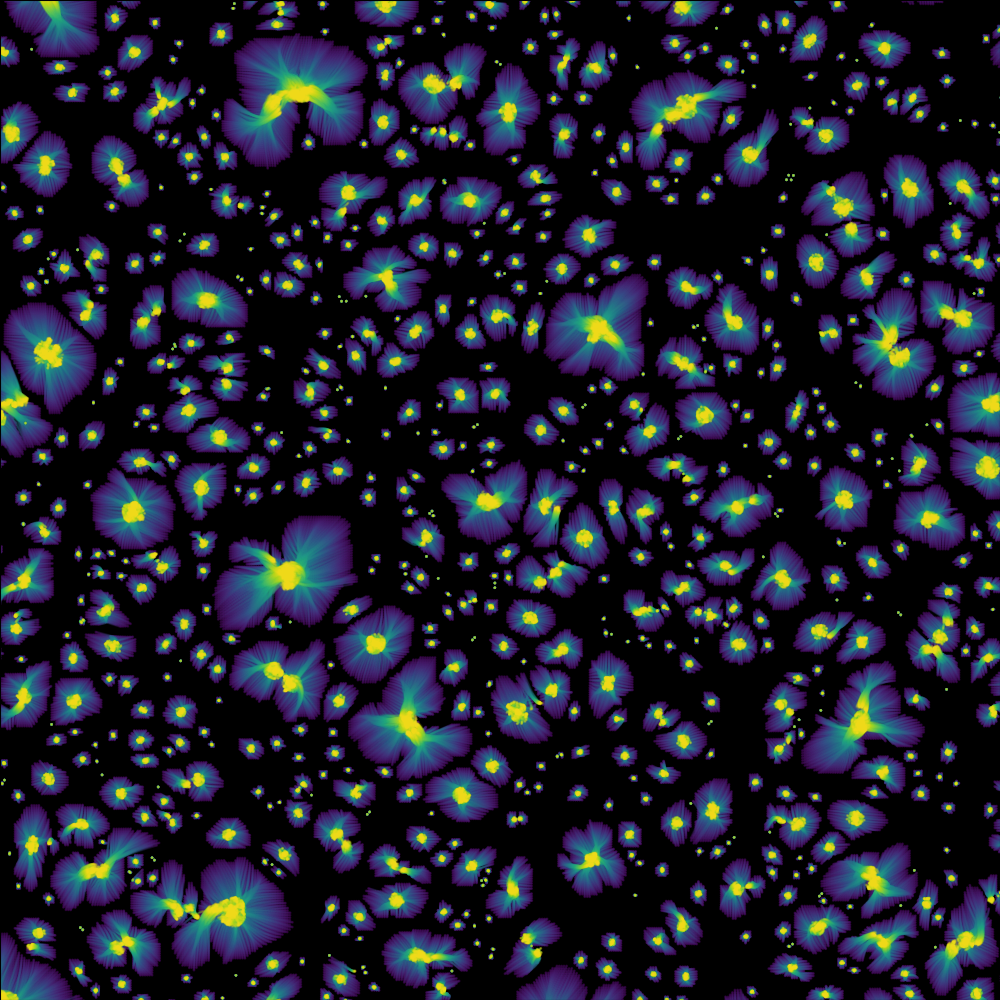}
    	\caption{
        Halo root systems in a 2D $N$-body simulation. At each timestep rendered, the particles are recentered to retain their Lagrangian (initial) centroids, removing the effect of bulk flows. Lagrangian outskirts of haloes, from the initial few snapshots, are in purple. Particles are rendered with increasing yellowness and decreasing opacity with time, until their final form appears in yellow.  Roots appear in greenish colors, rendered at intermediate times. The figure is 40 Mpc/$h$ on a side, drawn from a 50 Mpc/$h$ periodic box. An animation of particle deposition with time is at \url{https://neyrinck.github.io/haloroots_2d.mp4}.
        \vspace{1mm}}
        \label{fig:haloroots2d}
    \end{figure*}

There are essential differences between a 2D and the actual 3D universe, but as an introduction, it can be easier and computationally faster to investigate a 2D version of a 3D process. Fig.\ \ref{fig:haloroots2d} shows halo root systems from a 2D $N$-body simulation with $512^2$ particles. We used the python code of \citet{Hidding2020}, with an Einstein-de Sitter expansion history, a box size of 50 Mpc/$h$, a power-law power spectrum with slope $n=-1/2$, and an initial power-spectrum cutoff at 0.2 Mpc/$h$.

For halo finding in the final snapshot, we used a simplification\footnote{We tagged particles which differed in their place in line (lines being rows and columns along the axes, and $45^\circ$-diagonals), comparing initial and final orderings. The {\scshape origami} morphology is the number of orthogonal axes in which a particle was tagged as out-of-order; a halo particle has crossed along two axes. This only required comparison of ordered to `argsort' arrays in each row and column, extremely fast. However, this simple method misses some crossings, because assuming that orderings become randomized in a collapsed structure, there is some chance of a particle happening to recover its initial place in line. So, there were holes of untagged particles in collapsed patches; we filled these minimally, with the scikit-image \citep{scikit-image} `morphology.remove\_small\_holes' function. We then defined haloes as contiguous sets of halo particles in Lagrangian coordinates; note that this contrasts with below findings about streaks originating far outside haloes.} of the {\scshape origami} \citep{FalckEtal2012} algorithm, entirely coded in python, which we make publicly available, at (\url{https://github.com/neyrinck/origami2d}) with minimal 2D simulation code from \citet{Hidding2020}. While other halo-finders such as Friends-of-Friends (FOF) could be straightforwardly ported to 2D, with some tuning of the linking length, this dynamical algorithm requires no such tuning.

The figure is essentially in initial-conditions Lagrangian coordinates; dark purple footprints show the arrangement of haloes in the initial conditions. In colors that grow green ($a=0.5$) and then yellow ($a=1$) with time, particles' Eulerian positions are then drawn onto those footprints at each snapshot of spacing $\Delta a=0.02$, with bulk motions subtracted off from each halo at each snapshot. Note that 3D root systems appearing later do not have this temporal color information; instead, the colorscale there simply indicates higher deposited density. We include the temporal coloring here, first, because this was straightforward to do in 2D with the matplotlib `scatter' function. We also wanted to demonstrate explicitly what happens with time, and wanted every root-system peak to be similarly yellow, which would have been difficult if simply showing a rendered density. Particles grow increasingly concentrated with time, so to see earlier epochs here, we drew particles with decreasing opacity in time; the opacity $\alpha=0.5(0.02/a)^{0.5}$, starting at 0.5 at the first snapshot and decreasing to 0.07 at the final snapshot, $a=1$. The 3D root systems below, without explicit temporal coloring, carry more information about earlier structure than here, where the final color becomes mostly painted over with green and yellow in the center.
    
\section{Method}
\label{sec:method}
We measured root systems from `final' (redshift $z=0$) FOF haloes in an IllustrisTNG simulation \citep{PillepichEtal2018,NelsonSpringelEtal2019,NelsonPillepichEtal2019}, to enable us eventually to study correlations between their properties and the plentiful measurements that are readily accessible in the IllustrisTNG database at \url{https://www.tng-project.org/}. For each of the 100 snapshots from initial to final conditions, we binned all particles that end up in a final halo into cells, keeping the halo or protohalo centroid in the center.

Our main root-system measurements are from the 540$^3$-particle TNG50-3-Dark simulation, with box length 35 Mpc/$h$, with the same initial fluctuations as in the TNG50-3 simulations with the TNG galaxy-formation model. It has a concordance set of cosmological parameters specified by \citet{PlanckCosmoParams2015}. The size of this simulation in particle number is modest, but our implementation of root-system rendering was rather time-consuming; matches must be found between particle IDs in particular haloes, and particles in all files at all snapshots; it took $\sim 2$ days to render 1000 halo root systems. To study resolution effects, we also rendered some haloes from TNG50-2-Dark, with 8$\times$ mass resolution; each halo indeed took $\sim8^2\times$ longer. 

We rendered the simulations in 1.5 Mpc/$h$ and 0.75 Mpc/$h$ boxes with 128$^3$ voxels. We deposited particles using the `nearest-grid-point' method, i.e.\ binning with a simple histogram, weighting each particle by half the cosmic time between adjacent snapshots. There were enough snapshots that there was visually negligible particle discreteness noise in this, but if fewer snapshots were available, it would make sense to draw lines between particles instead of just depositing their mass.

\section{Results}
\label{sec:results}
\begin{figure*}
        \centering
    	\includegraphics[width=\textwidth]{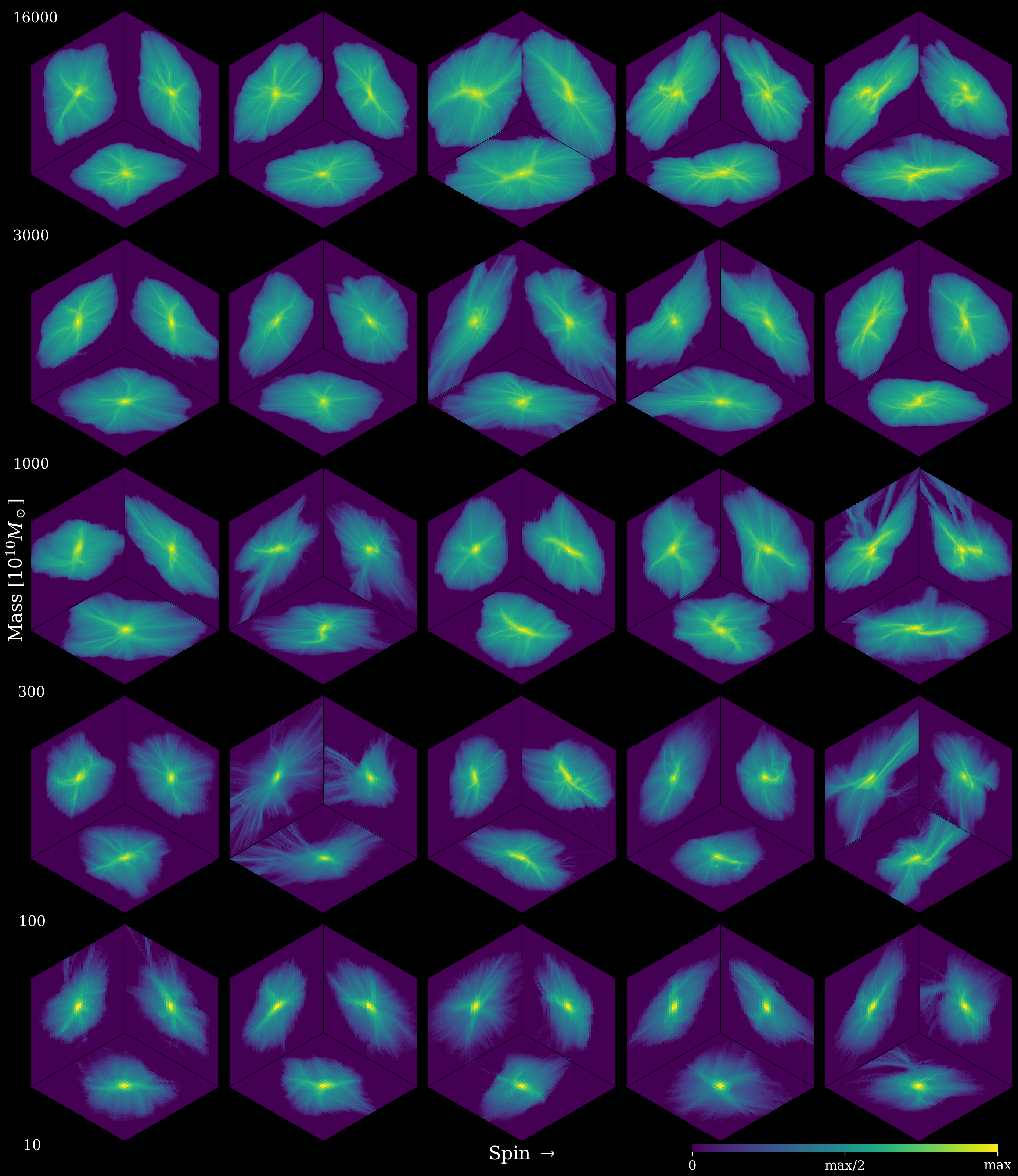}
    	\caption{Projected halo root systems (HRS) spanning ranges of mass and final halo spin. Each hexagonal panel shows the log-density of 3 projections along the cardinal axes of the rendered cube (note that the color in Fig.\ \ref{fig:haloroots2d}, in contrast, shows deposition time; it is interesting that these end up looking similar). The cube is scaled to be of side length $0.8V_{\rm HRS}^{1/3}$, where $V_{\rm HRS}$, roughly proportional to the mass, is the number of nonzero-density voxels in the HRS. Deep purple indicates 0; bright yellow indicates the maximum, typically $\sim 2\times10^5$ deposited particles (bottom row), up to $\sim 1.2\times 10^6$ (top row).
        Mass bin edges (in units of $10^{10} M_\odot$) of each row are shown at left. The top row shows galaxy-cluster-size haloes, containing hundreds of galaxies each, according to the TNG model. By the second-to-last and last rows, some haloes contain only a single galaxy. In columns, haloes in that bin are shown as close as possible to the minimum, 25th percentile, median, 75th percentile, and maximum spin.\vspace{1mm}}
        \label{fig:mass_spin}
    \end{figure*}

We show projected root systems for 25 haloes in Fig.\ \ref{fig:mass_spin}, spanning ranges of mass and halo spin. Each row represents a mass bin; within that, columns show haloes of increasing final spin in the mass bin, from minimum to maximum. We measured `spin' as the magnitude of the total angular momentum, summed with a cross product of $z=0$ particle displacements and velocities away from the halo's position and velocity centroids. For this simple but not dimensionless spin parameter $L$, we observed a trend $L\propto M^{\sim1.5}$, roughly that seen previously \citep[e.g.][]{HeavensPeacock1988,NeyrinckEtal2020}; to eliminate bias from this trend within a mass bin, we actually ranked the quantity $(L/M^{1.5})$.

The outer boundaries of the root systems are essentially their Lagrangian protohaloes, i.e.\ the patch of initial-conditions uniform-density space that collapses to form the halo. This is because, in comoving coordinates, infalling particles move toward the centroid; new voxels are not added to the root system if, for each snapshot, each particle continues to occupy voxels previously occupied by some particle. This is not always perfectly true, though; protohaloes may consist of multiple connected components, producing `streaks' related to the analysis in \S\ref{sec:streaks}.

Animations of root systems' rotating 3D structure can be found at \url{https://neyrinck.github.io/halo_gif_table.html}, showing the 100 haloes with the highest mass, \url{https://neyrinck.github.io/halo_gif_table_skip10.html}, showing every 10th of the 1000 highest-mass haloes, and \url{https://neyrinck.github.io/halo_gif_spin_table.html}, showing the same grid as appears in Fig.\ \ref{fig:mass_spin}. For these, we rendered the haloes with 10 equally-spaced contours of the log-density up to the maximum, with increasing opacity.

A first finding is the surprising (to us) prevalence of streaks reaching far outside of a convex shape. We emphasize though that we expect this to depend on the halo finder and definition; we used haloes defined by a standard FOF method. In this, particles within a physically motivated linking length of each other are grouped together as a halo. This is a conveniently simple definition, but it does not enforce that the haloes are physically persistent through time; it does not, in particular, include an `unbinding' step which removes particles not gravitationally bound to the halo. Although we could have readily used another halo definition for our analysis \citep[see, e.g.,][]{KnebeEtal2011}, we used FOF haloes since they were the most readily available in the TNG database, and also underlie most halo quantities in it. These streaks are further discussed in Sec. \ref{sec:streaks}.

\subsection{Root-system complexity}
\label{sec:roots}
We define a {\it root} as an accumulation of pathlines in the root system that appears roughly as a cosmic-web filament (a nearly 1-dimensional curve). There could be some 2D equivalents of cosmic walls carrying in matter, either genuine walls that collapse lengthwise into the halo, or filaments perpendicular to the infall that sweep out a 2D manifold. But we found them not evident visually, so we focus on the filament-like roots. Because it is ambiguous how physically relevant the streaks are that poke out of the central protohalo (the nearly convex shape filled with particles initially), we concentrate our attention on roots inside that central protohalo.

A quick visual impression finds more `structure' (more roots, and more branching) in a high-mass halo than a low-mass one. This suggests a correspondence between filaments as they exist at a given epoch and previous mass accretion, since the number of filaments around a galaxy or density peak generally increases with mass too, which has been found analytically \citep{CodisEtal2018}, in simulations \citep{KraljicEtal2020}, and in observations \citep{GalarragaEspinosaEtal2023}. There have been similar findings for galaxy clusters, in simulations \citep{aragon2010} and observations \citep{Euclid2025}. 

Also, based just on a visual impression, we find that the roots in a high-spin root system are generally more curved, or tortuous (a word used to describe a curvy river system, not to be confused with `torturous'). Some even have a gap at the center between two blobs. This could happen if the blobs are linked in the FOF method, but have not yet physically merged. This is entirely expected; although halo spin may ultimately be sourced by extended regions of the initial velocity field \citep{NeyrinckEtal2020}, an ultimately spinning structure may often manifest as a merger of haloes in spiraling orbits.

However, although most of this paper is about the roots, the most visually striking aspects of the root systems, there is a substantial background of matter in them, as well. A fair fraction of matter seems to come straight into the halo, without noticeably clumping.

It would be interesting to undertake a comprehensive study of the abundance, structure, branching, and fractal dimensions of halo roots, as has occurred in other disciplines. Many branching structures have objective boundaries, such as cell walls for neurons, blood-vessel boundaries for the circulatory system, bronchial boundaries in the lungs, road edges for transportation, or river banks for river networks. Given their clear medical value, blood-vessel analysis algorithms \citep[e.g.\ the review][]{moccia2018blood} have matured for decades, and even underlie some common cosmic-web analysis techniques \citep{aragonMMF,CautunEtal2013,CautunEtal2014}.

But a difficulty is that there is no easy ground truth of what is a root and a root boundary within a protohalo. In the case of the usual cosmic web, dynamical particle-crossing information helps to objectively define some structures \citep{FalckEtal2012}. Similarly, we imagine that tagging particle-trajectory crossings could define roots objectively and decisively. Referencing the origami view of cosmic-web formation \citep{Neyrinck6OSME2015}, because the root systems show the infall of all particles, the roots are essentially creases in a fabric, where an initially uniform, uncreased 3D patch of the universe bunches together. But we leave this objective root detection possibility for future follow-up, because this particle-crossing would be difficult to measure with the glass arrangement of particles in the TNG database. 

The galaxy root system built from pathlines of gas that ultimately forms stars \citep{AragonCalvoIsochrones}, we expect, would have more objective boundaries, as well. In that paper, infalling gas, colored according to the time that gas ultimately would form stars in a galaxy, seems to form a branching structure; see the upper-right panel of its Fig.\ 1. Unfortunately, in IllustrisTNG, analyzing gas infall would be much more difficult than dark matter infall, since in Arepo \citep{Arepo2020}, gas particles are tracers, not representing the same matter throughout the simulation. They do not precisely go with the flow, and gas can flow through gas-cell walls. It should be possible to track this star-forming gas using tracers \citep[a paper addressing this sort of thing is][]{GenelEtal2013}, but for now we just look at the dark matter. The dark matter root system also does not depend on the whims of simulated random numbers involved in activating star formation.

\begin{figure*}
        \centering
    	\includegraphics[width=\textwidth]{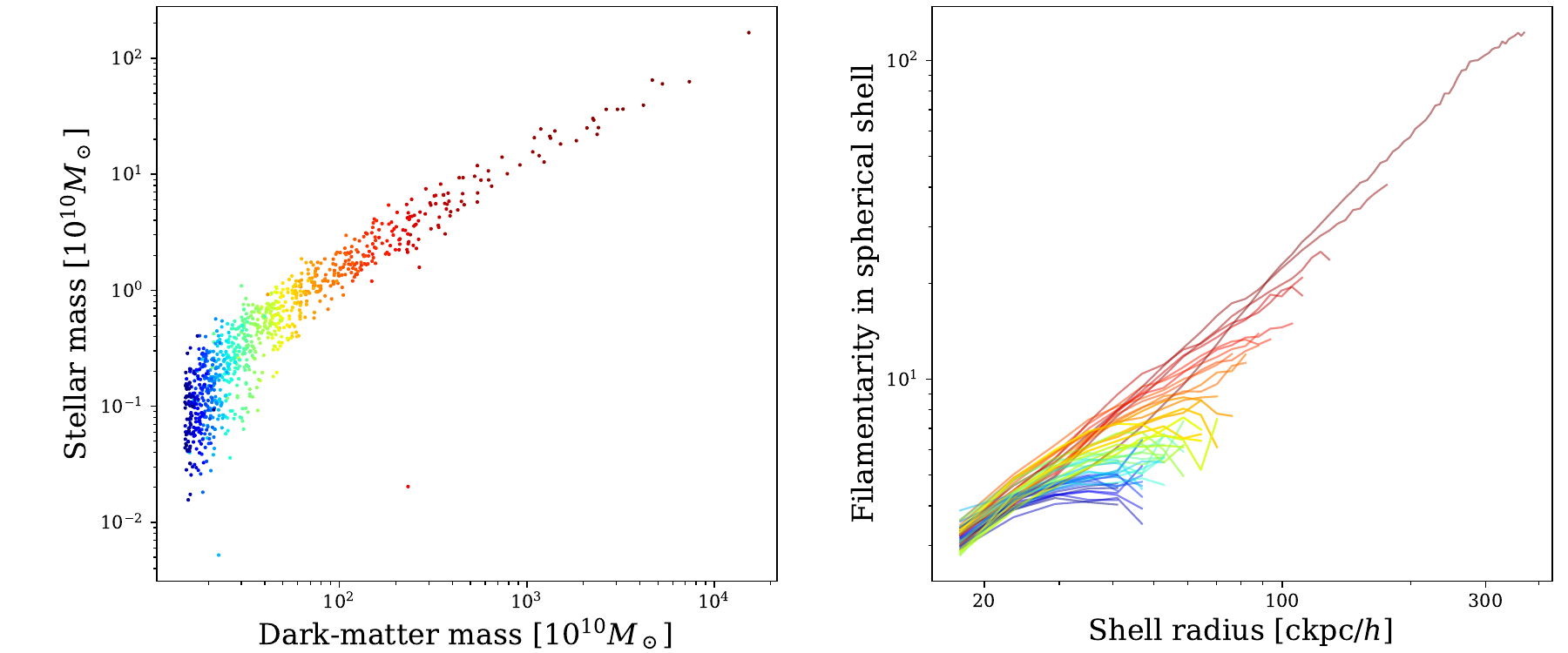}
    	\caption{\textbf{\textit{Left}}: Scatter plot of dark-matter and stellar masses for the 1000 most massive haloes, as reported in the TNG50 database. Dots are colored in a rainbow colorscale, linearly increasing with the rank of halo mass.\\
        \textbf{\textit{Right}}: Halo filamentarity as a function of distance to the halo center. Each curve is an average of measurements for 20 haloes in a mass bin, and colored the same way as at left. `Filamentarity' here means the blobbiness- and density-weighted number of filaments found by a 2D Hessian-based blob finder, described in the Appendix. The lowest radius shown corresponds to only 3 pixels, and the upper radius to the half-radius of the box, 63 pixels. As halo mass increases, the filamentarity generally increases at a given radius and there is a reduced filamentary downturn at the edge of the halo root system. This is consistent with the visual impression that the halo root systems do not have similar structure across mass; the number and branching of filaments increases with mass. Here, `ckpc' means comoving kiloparsec, used to remind the reader that the root systems are projected in comoving coordinates.}
        \label{fig:filamentarity_mass}
    \end{figure*}

As a first quantitative measurement of the root systems, we use a multiscale `blob' finder (where a blob is roughly circular) on spherical shells in voxel-length radius increments from the root-system centroid. Instead of making a decisive cut for each blob as to whether it counts as a root or not, we define a `filamentarity' as a probabilistic field that quantifies blobbiness at many radii. We would like to detect clear branching events as well in the root system, but this probabilistic summing makes that more challenging. The Appendix contains details of this procedure, and Fig.\ \ref{fig:filamentarity_mass} shows the results.

At fixed shell radius, from $r\sim 30$-80 ckpc/$h$, the filamentarity generally increases with halo mass. The filamentarity also generally increases with radius for a given curve (corresponding to a single 20-halo mass bin). The slope of the upper limit of the curves is filamentarity $f\propto r^1$, which is substantial, but is less steep than the surface area of the shell ($\propto r^2$); this is the expected rise if the blob-finding were due to noise. The rise suggests branching, since new roots arise at larger radius. This rise persists throughout the radius range of each curve for high-mass haloes, but it turns over for lower-mass haloes, indicating less branching in these root systems.

\subsection{Root systems and cosmic-web detachment}
\label{sec:cwd}
Fig.\ \ref{fig:filamentarity_mass} also shows a plot of the dark-matter and stellar masses for these haloes, as reported in the TNG database. We did this to investigate the possible relation of root systems to the cosmic-web detachment \citep[CWD;][]{AragonCalvoEtal2019} mechanism for external star-formation quenching that some of us proposed. `Quenching' refers to a drastic, possibly sudden reduction in the star-formation rate. High-mass galaxies and clusters are often quenched; in Fig.\ \ref{fig:filamentarity_mass}, this appears as a reduction in the slope of the stellar mass to dark-matter-halo mass relation at high mass.

The CWD mechanism is as follows: when haloes first form (from infall rather than mergers), they are generally surrounded by primordial filaments, along which gas and dark matter enter the halo, ultimately forming stars. These are halo roots in the present picture. The `web detachment' occurs when these roots get `detached,' or disrupted, which may happen through a merger with another halo, or collision with a filament. In the CWD paper, we quantified these detachment events with stream crossing on larger scales than the primordial haloes.

Because the halo root system is a single 3D representation of a halo merger tree, these CWD events should often show up in the root system as major root branchings. Our measurements indicate increased root-system filamentarity and branching for the most massive haloes, i.e., more merging and CWD in these haloes, disrupting star formation in purely gravitational dynamics without reference to gas.

However, the star-formation picture in the TNG model includes many processes internal to galaxies that are explicitly stochastic; neither in the simulation nor in nature would we expect CWD to completely determine, or perhaps even be the primary cause of star-formation quenching. Indeed, any simulation that aims to produce galaxy observables from a fixed set of initial conditions is best viewed as a particular realization of a probabilistic ensemble \citep{GenelEtal2019,NeyrinckEtal2022}. Still, we expect some external mechanisms clearly predictable from the initial conditions, such as CWD, to show up somehow in halo root systems. In haloes containing many galaxies, random virial motions would mechanically inhibit star-forming gas from finding its way into galaxies. But other mechanisms could dominate quenching at lower mass without indications in the halo or galaxy root system, mechanisms such as internally triggered AGN. 

\subsection{Infall anisotropy}
\label{sec:anisotropy}
Because root systems encapsulate the formation history of a halo, they could be useful for studying issues such as possible anisotropies that can occur with a small box, such as the 50 Mpc/$h$ box we use.

\begin{figure}
    \includegraphics[width=0.8\columnwidth]{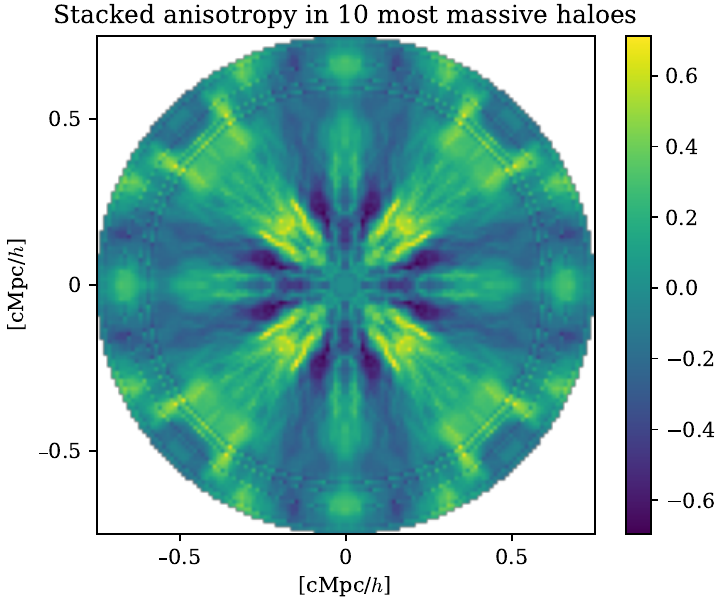}
    \includegraphics[width=0.8\columnwidth]{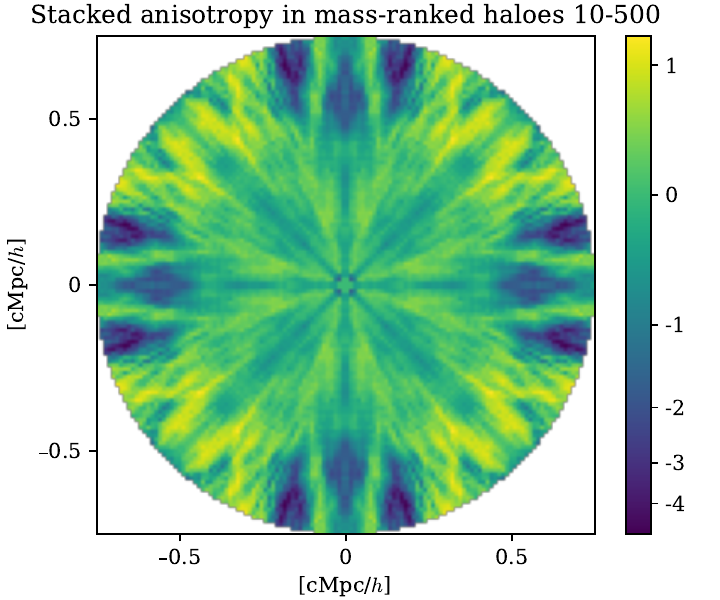}
    \caption{Heatmaps showing anistotropy in two halo bins: (\textbf{\textit{Top}}) the most-massive 10 haloes in the sample, and (\textbf{\textit{Bottom}}) the rest of the 500 most-massive haloes. Heatmap values show the relative prevalence or absence of paths through that pixel, in units of the noise level in the stack. Substantial anisotropy arises in outer regions. Only root systems that reach the radial extent of a pixel are counted in the stack there. Only 53 root systems contribute to the stack at the edge (and 450 in the middle), but this is taken into account in the noise estimate.}
    \label{fig:anisotropy}
\end{figure}

Any single root system is likely to have high anisotropy, but a stack of many of them could reveal if, for example, infall is preferentially along the Cartesian axes, or diagonal to them at certain angles. Fig.\ \ref{fig:anisotropy} shows this anisotropy for two groups of haloes: first, the 10 most massive haloes, whose root systems typically occupy most of the aperture shown, and the rest of the 500 most massive haloes, many of whose root systems are much smaller.

The figures show angular fluctuations in a symmetrized $\log(1+\rho_{\rm HRS,\ projected,\ stacked})$. To get a symmetrized angular 2D contribution from each root system, we project it along the $x$, $y$, and $z$ axes, stack those, and then symmetrize the stack over reflections and transpositions. This gives a total contribution of 24 2D images per root system; the total has independent information only between angles $0$ and $45^\circ$. The heatmap value in a pixel is the mean $\log(1+\rho_{\rm HRS,\ projected,\ stacked})$ for root systems extending to pixels at that radius. (HRS stands for `halo root systems'.) We subtract off an interpolated radial average in each bin, to show only angular fluctuations. This appears in units of the error, i.e., the standard deviation among haloes divided by $\sqrt{N_{\rm HRS}(r)-1}$, the number of haloes with roots extending, in whatever direction, to the radial distance to that pixel.

The first, high-mass stack appears to have substantial anisotropies along the diagonals at $r\approx 0.2$ cMpc/$h$. This comes from only 10 haloes, but these most massive haloes occupy the largest Lagrangian volumes, and are where we would most imagine the effects of the low box size to show up.

The second, lower-mass stack also shows greater accretion along diagonal directions, at large distances from the halo center, $r>0.5$ cMpc/$h$. There are relatively few root systems involved in this high-$r$ anisotropy; the number in this mass range that reach the edge is only 52 out of 490. Still, it seems to be statistically significant, reaching a signal-to-noise level of several (note the nonlinear stretch of the bottom colorscale, done using a sinh transform).

There seems to be compelling anisotropy in the stacked root systems, which we show mainly to demonstrate their use for this application. This is a reminder of how a 50 Mpc/$h$ box is small by cosmological standards \citep[e.g.][]{RaczEtal2021}, possibly affecting galaxies in the simulation that are naively much smaller than the 1.5 Mpc/$h$ window we use. Still, we do not mean to advise against the use of the TNG simulations for any particular purpose; even if that purpose would require strict isotropy in halo assembly, further analysis would be necessary to establish that there is a problem.

\subsection{Root-system shape}
\label{sec:streaks}
Above, around Fig.\ \ref{fig:mass_spin}, we noted that many root systems have `streaks' extending rather far from their convex cores. Thus, at least with a FOF halo definition, an assumption that a protohalo is a convex shape seems to be poor; here, we examine this picture in more detail. We base all measurements in this section on spheres of radius 0.75 cMpc/$h$ that inscribe root-system datacubes, even though some of them extend outside that sphere.

We also observed that most Lagrangian protohaloes were not even single blobs topologically; in the protohalo (the particles in the FOF halo, mapped back to the initial snapshot), we often observed a layer inside the outer boundary that had particles missing. We guess that the outermost particles in the protohalo are on their way through its center at $z=0$, undergoing first passage, but some particles inside that layer in the initial conditions are splashback particles at $z=0$. This is the halo's least-dense region, farthest from the center; if any halo particles would fail to meet the FOF linking-length criterion, it would be these.

\begin{figure*}
    \centering
    \includegraphics[width=2\columnwidth]{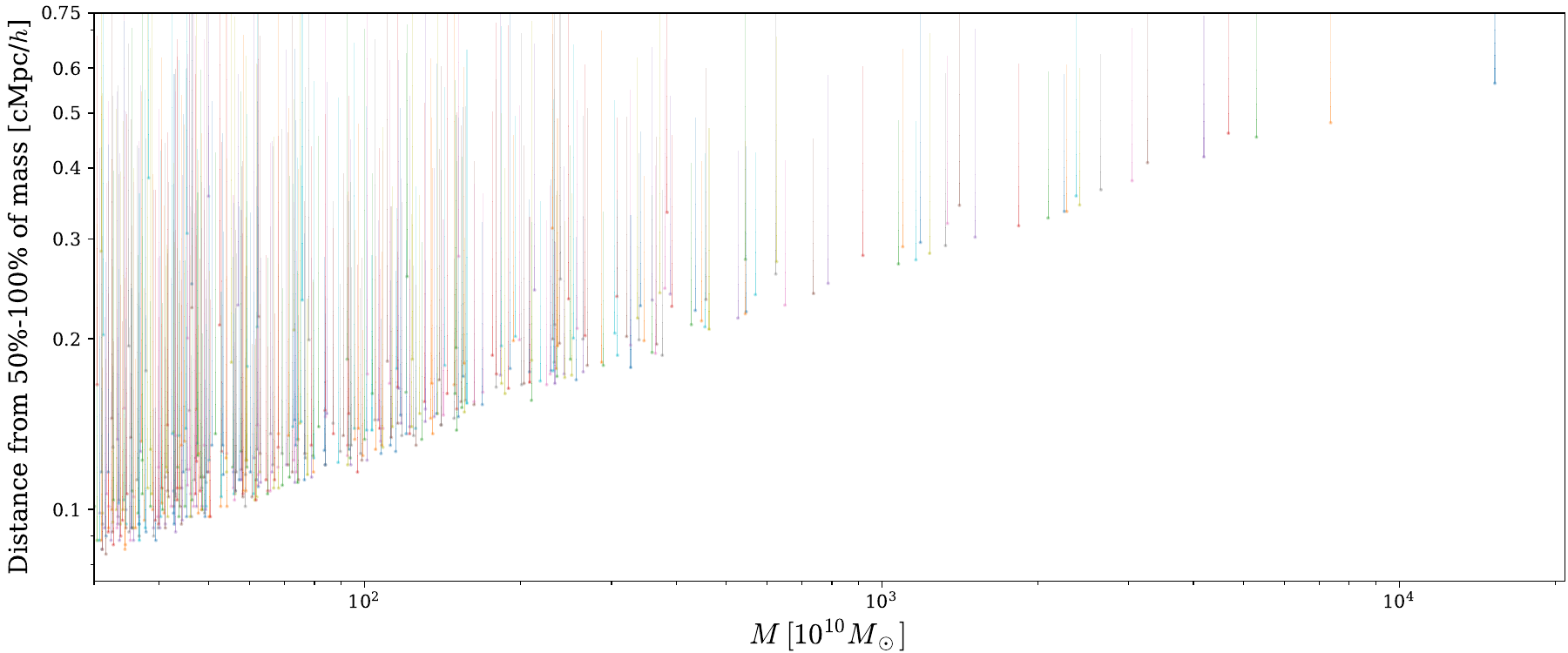}
    \includegraphics[width=2\columnwidth]{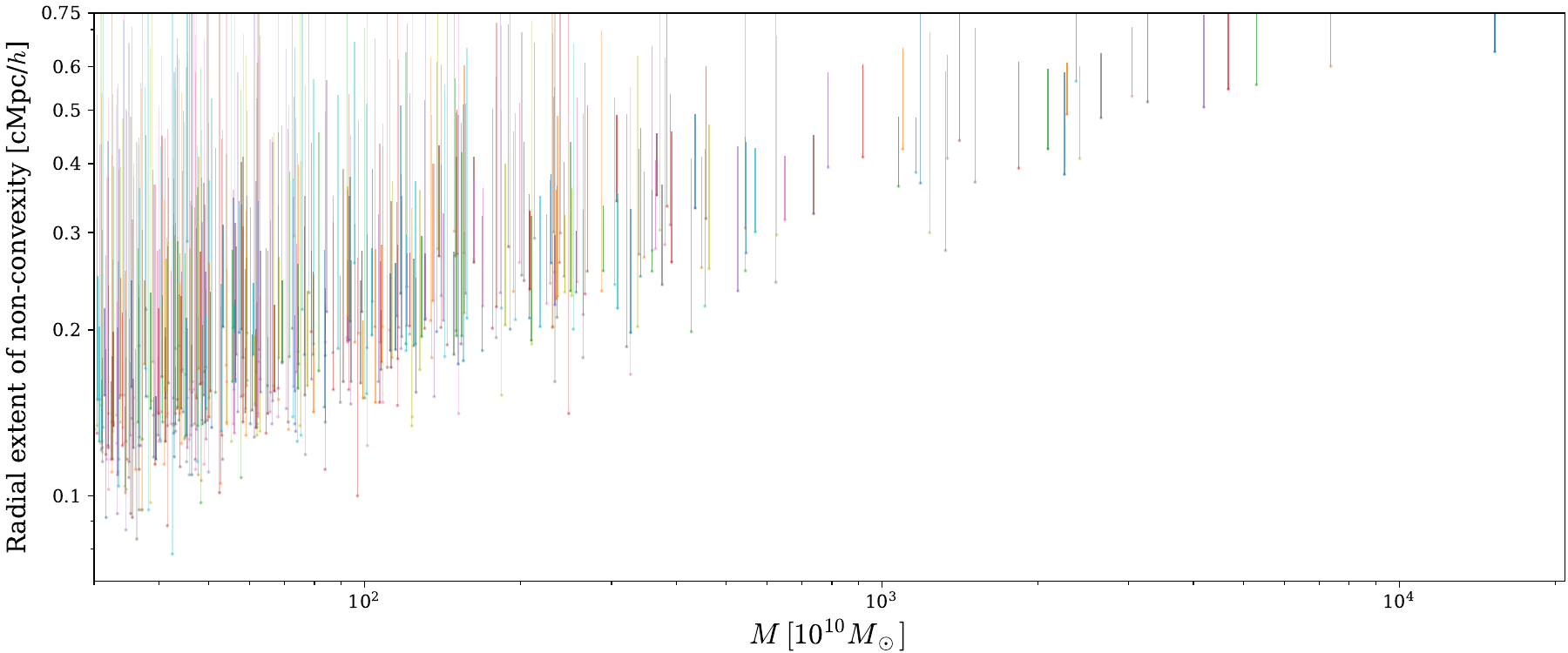}    
    \caption{{\textbf{\textit{Top}}}: The radial extent of the outer layers of halo root systems, shown with decreasing opacity going from each root system's half-volume radius (50\%, receiving a dot as well) to its farthest extent in a 0.75 Mpc/$h$ sphere (100\%).\\
    {\textbf{\textit{Bottom}}}: The radial extent of `non-convexity' of the root systems. For each root system, lines are drawn from the farthest extent of its convex core (see text for definition), receiving a dot, to the farthest extent overall.}
    \label{fig:streakiness}
\end{figure*}

In Fig.\ \ref{fig:streakiness} we show the results of root-system measurements sensitive to these streaks. In the top panel, we draw lines from the half-volume radius of the root system (receiving a small dot as well) to its farthest radial extent.  The segments starting from the radii enclosing 50\%, 60\%, 70\%, 80\%, and 90\% of the volume have opacity 0.5, 0.4, 0.3, 0.25, and 0.2.

In the bottom panel, we try to show the radial extent of each root system's `non-convexity.' We draw lines from the farthest radial extent of the `maximal convex core' (receiving a small dot) to the farthest radial extent of the root system. To show more compact root systems more clearly, we give a line greater opacity and thickness when the difference between these radii for a root system is small. The `maximal convex core' was estimated using a star-shaped convex expansion from a seed. The steps in this process are (1) Remove any `small holes' in the root system with the scikit-image `morphology.remove\_small\_holes' function. (In most cases, this made a small difference.) (2) Start at a `seed,' the center of the largest sphere entirely within the root system. (3) Cast many (256) rays in nearly uniform directions. (4) For each direction, find the distance to the root-system boundary. (5) Build a candidate polyhedron with vertices at those ray endpoints. (6) Shrink offending vertices/rays until their convex hull is fully inside the mask.

Generally, this figure shows that larger haloes are more convex and compact than smaller haloes; smaller haloes are the ones that can have `streaks' reaching far outside their convex cores. This makes some sense, as the largest haloes have radius of order the largest comoving displacements in the simulation, but if a patch enters a smaller halo with a similarly large displacement, that streak would be large compared to the halo size.

Why are there such streaks? For this to happen, particles originating far away must enter the halo, but nearer particles along or near the streak strangely appear not to enter. Perhaps the faraway particles participated in few-body interactions that took them into the halo, bypassing the intervening particles. Perhaps these are particles that were once in one halo, but swapped into another inside a structure like a filament. As we see below in Fig.\ \ref{fig:resolution}, these streaks largely persist at higher resolution, and so seem to be physically real. Haloes can be defined as hole-free regions of Lagrangian space (in 2D, shown in Fig.\ \ref{fig:haloroots2d}) but maybe that definition has physical (instead of numerical) exceptions that these results illustrate. Indeed, many haloes are visibly non-convex in Fig.\ \ref{fig:haloroots2d}, surprisingly so to us even given that they are often lumpy and `peakless' \citep{LudlowPorciani2011}.

\subsection{Resolution effects}
\label{sec:resolution}
An effect that could artificially mimic our claimed reduced number and branching of roots in lower-mass halo root systems is if we were smearing over structure at lower mass. This could either come from too-low simulation mass resolution (too few particles per halo), or from fixed-size voxels that are too big. The number of particles in the lowest-mass halo we investigate is 6739, which seemed adequate to us, but it is still worth checking what would happen with increased resolution.

\begin{figure*}
        \centering
            \includegraphics[scale=0.99]{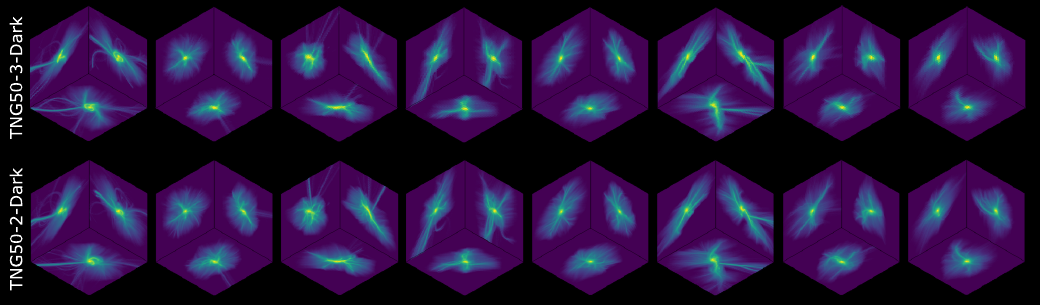}
    	\caption{Comparison between projected root systems, as in Fig.\ \ref{fig:mass_spin}, for 8 haloes in TNG50-3-Dark (top row), and the same haloes with $8\times$ higher mass resolution in TNG50-2-Dark, with twice as many voxels in each dimension (bottom row). They are the 250th, 350th ... to the 950th most massive in TNG50-2-Dark, and range in mass from 1.4$\times10^{11}$ to 8.3$\times10^{11}M_\odot$. Haloes in each pair look quite similar, but there are small differences.\vspace{3mm}}
        \label{fig:resolution}
    \end{figure*}

To study whether this is the case, in Fig.\ \ref{fig:resolution} we show 8 halo root systems from our fiducial TNG50-3-Dark simulation, each side-by-side with halo root systems with 8 times more particles.

We note also that it is not precisely accurate to call these the `same haloes' simulated with different resolutions, because the haloes in the higher-resolution simulation are not only built in the initial conditions from 8$\times$ the particles, but also have additional small-scale structure from $8\times$ the Fourier modes. Given this, we find the changes in the root system to be remarkably small; the reason for this could be the small amplitude of fluctuations contributed by these added modes, i.e.\ because the power spectrum is steeply declining with wavenumber in this regime.

Another not-entirely-expected finding from this crude resolution study is that the externally arriving streaks that occur in many of these halo root systems are generally present (with some fluctuations in their density) in both simulations, supporting the view that they are physical instead of numerical.

The appearance of each projected halo seems to have modest dependence on resolution, although there are a few exceptions visible in Fig.\ \ref{fig:resolution}. We conclude that the trend of fewer roots for lower-mass haloes seen in Fig.\ \ref{fig:filamentarity_mass} survives consideration of resolution effects.

\section{Galaxy information}
Root systems give a visual, geometric representation of the accretion, infall, or merger history in a halo (or galaxy). Thus, we assert that they are useful summaries of the influence that the assembly history and the density field within the protohalo could have had on the final spatial arrangement and properties of the halo. The protohalo density field contains this information too, but not in a way as visually accessible to humans whose eyes are sensitive to branching systems. This patch of all the matter that, through physical flows, could have causally affected a halo and galaxies within it has been called the {\it matter horizon} \citep{EllisStoeger2009}. This horizon is much nearer than cosmological event or particle horizons, but it dominates the causal influence among parts of the light cone.

One of us suggested that the initial density field within the Lagrangian patch that collapsed to form a halo quantifies the information available to form it \citep{Neyrinck2015}. In principle, the initial velocity field within that patch should carry similar information as in halo root system; the root system is constructed by carrying the initial velocity field forward. But we find the root system to be particularly visually comprehensible, and to display features such as mergers directly.

As we commented in \S \ref{sec:resolution}, small-scale velocity modes have little influence on the root system, but they would ultimately have some influence on the detailed dynamics, because of the high degree of chaos in a halo; they might ultimately reach halo scale. In principle, the root system, perhaps with a bit further information about the tidal field, should directly give the final structure of the halo. But the root system is numerically stable and informative, a semi-Lagrangian picture that depends on flows that should be laminar on scales larger than sub-clumps \citep{NeyrinckEtal2022}. The detailed Eulerian final structure, on the other hand, is quite sensitive to initial conditions.

\section{Conclusion}
\label{sec:conclusion}
The cosmic web of nodes, filaments, walls and voids is a well-known branching network in cosmology, with similarities to other branching systems such as river networks; for a popular-level article about these correspondences, see \citet{Neyrinck2025CosmicWebEarth}. Unraveling a collapsed halo in time reveals the correspondence to some of these systems inside a halo or node, not just in the structure around it.

A prime example of an analog system is a river network leading to a sea. Fig.\ \ref{fig:haloroots2d} has parallels to a deluge of water falling on a landscape at the initial conditions, with the gravitational potential playing the role of the initial altitude on the landscape. As matter coalesces gravitationally, it etches valleys (filaments) into the potential, some of them branching, and each basin settles down to form an inland sea. As time progresses, each root system continues to develop, as matter continues to accrete onto it.

We imagine that each root is typically built from the infall of a satellite halo into a central blob, but it may also have substantial contribution from a filament falling in lengthwise. The finding that higher-mass haloes have more abundant and branching roots is consistent with both possibilities.

A high-mass halo would typically contain a galaxy cluster which may contain hundreds of galaxies; if each of these contributes a root, some of them merging, this would account for the increased root prevalence at high mass. If, alternatively, we associate roots directly with filaments, increased root prevalence at high mass accords with findings that higher-mass haloes are connected to more filaments; for example, \citet{CodisEtal2018} found this even analytically, using the theory of density peaks in a Gaussian random field. Conceptually, this is because high-mass haloes are larger peaks, with larger catchment basins, and therefore more branches in the flow.

The cosmic-web context of the haloes is also important to consider; the largest, cluster-size haloes occupy the largest nodes of the cosmic web. Lower-mass haloes, on the other hand, can occupy a variety of cosmic-web environments. If they are in a large void, they may have rather undisturbed accretion along scaled-down filaments. But if they are in a filament or wall (i.e.\ their primordial webs are `detached'), their dynamics become more enmeshed with their surroundings. Because filaments form in particular tidal environments that persist throughout cosmic time, the roots of a particular halo are likely to align with filaments outside the halo. Looking at how root systems depend on cosmic-web environment is another avenue for further study.

The way a halo assembles relates to the problem of optimal transport \citep[see the recent data-science text,][]{PeyreOptimalTransportBook}. This was originally developed to study efficient resource transportation and allocation, but it has been fruitfully applied in cosmology \citep{FrischEtal2002,MohayaeeEtal2006,LevyEtal2021} to provide a simple way to infer the initial-conditions arrangement of matter that is presently clumped. In solutions of optimal transport, it has been found that branching structures can emerge and play a key role if an increase in flux through a certain patch increases a conductivity parameter through a medium \citep{facca2021branching}. Blindly interpreting that cosmologically, perhaps a reason that an initially uniform protohalo crumples through roots as it collapses to form a halo is for efficient gravitational transport.

\section*{Acknowledgements}
For this paper, our only use of generative AI was to flag a few grammatical typos in final proofreading. We thank the developers and maintainers of the IllustrisTNG database that enabled this work. MN thanks the organizers of the conferences `Mind the Gap: Galaxies and the Large-Scale Structure' and `Cosmic Flows 2025' for providing stimulating environments that encouraged the ideas and work presented here, and thanks Antonela Taverna for providing an in-preparation python implementation of the MMF. MAC acknowledges support from the Programa de Apoyo a Proyectos de Investigación e Innovación Tecnológica (PAPIIT) IN115224 (also enabling a visit by MN) and CONAHCyT project CF-2023-I-1971. IS and MN acknowledge NASA ROSES grants 80NSSC24K1489 and 24-ADAP24-0074, and contract number 80NM0018F0610 via a JPL sub-award.

\section*{Data Availability}
The main results were rendered using data available through the IllustrisTNG database at \url{https://www.tng-project.org/}, making use of the servers there. Codes used for that, and other results, can be shared upon reasonable request to the authors. We are releasing one code developed for this paper publicly, a 2D halo-finder, at \url{https://github.com/neyrinck/origami2d}.

\bibliographystyle{mnras}

\bibliography{refs,refs-spiderweb}


\appendix
\section{Blob-finder details}
\label{ap:blobfinder}
To quantify root abundance in the root system, we count blobs using a 2D Hessian-based blob filter, applied to a spherical shell of radius $r$ around the halo root system centroid. It is a similar method to one introduced to bring out multiscale features resembling blood vessels by \citet{frangi1998multiscale}, but instead is tuned to find (smeared) pointlike blobs; it is the node-finding part of \citet{Aragon-CalvoEtal2010}'s Multiscale Morphology Filter (MMF).

The Hessian-based blob filter works on a field $\rho$ smoothed on a scale $s$, $G(\rho,s)$. At a single scale $s$, it returns the magnitude of the smallest eigenvalue of the Hessian matrix of the image, where that eigenvalue is negative; otherwise, zero:
\begin{equation}
    B(\rho,s)=|e_0(G(\rho,s))|(e_0(\rho,s) < 0),
\end{equation}
where $e_0$ is the smallest eigenvalue of the Hessian of the field. The final multiscale filter is the maximum of this field over several measured scales.

We used the simple equirectangular projection to translate right ascension and declination on the shell onto a grid, padding the edges substantially to avoid edge effects, and using only declination between -60 and 60 because of the distortion at the poles. To cover the rest of the spherical shell, we measured $B$ from two other rotations of the halo root system, and took the median of the average blobbiness fields among the three rotations.

\begin{figure}
        \centering
    	\includegraphics[width=\columnwidth]{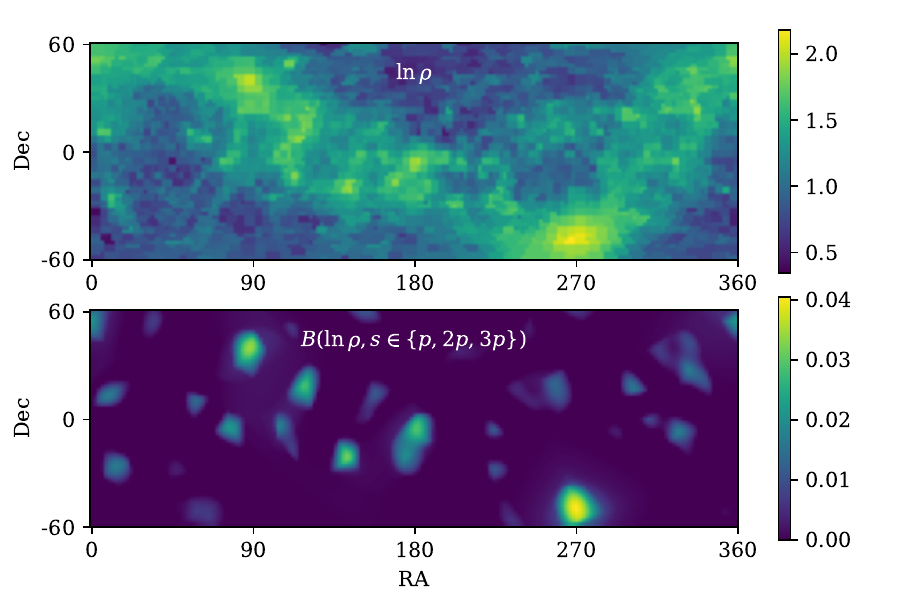}
    	\caption{An example `blobbiness' field, measured from the log-density in a spherical shell of a halo root system. We looked for blobs at three scales (1, 2, and 3 times the scale size in degrees $p$ of a pixel at radius $r$), and combined them into a multiscale-blob field.}
        \label{fig:blobbiness}
    \end{figure}

Fig.\ \ref{fig:blobbiness} shows a spherical shell through the cubic grid, for halo 40, and at a radius 15 voxels away from the center, and the multiscale `blobbiness' field $B$. The pixelation of the grid, and its distortion from the spherical projection, is visible at top. The smoothing scales we used in the filter were 1, 2, and 3 times the pixel scale, i.e. the angular size of a pixel a distance $r$ away from the center, at the equator.

There is substantial room for improvement in this method to identify infall roots, but we would like to think this is adequate for a first quantification of them. Ultimately, we would like to map the roots as has been done in other vascular systems. Some discussion of the challenges and further ideas for quantifying them is in the main text.

\label{lastpage}
\end{document}